# Analysis of bi-material interface cracks with complex weighting functions and non-standard quadrature

Ali R. Hadjesfandiari, Gary F. Dargush

*Department of Mechanical and Aerospace Engineering*
*University at Buffalo, State University of New York*
*Buffalo, NY 14260 USA*
ah@buffalo.edu, gdargush@buffalo.edu

**Abstract**

A boundary element formulation is developed to determine the complex stress intensity factors associated with cracks on the interface between dissimilar materials. This represents an extension of the methodology developed previously by the authors for determination of free-edge generalized stress intensity factors on bi-material interfaces, which employs displacements and weighted tractions as primary variables. However, in the present work, the characteristic oscillating stress singularity is addressed through the introduction of complex weighting functions for both displacements and tractions, along with corresponding non-standard numerical quadrature formulas. As a result, this boundary-only approach provides extremely accurate mesh-independent solutions for a range of two-dimensional interface crack problems. A number of computational examples are considered to assess the performance of the method in comparison with analytical solutions and previous work on the subject. As a final application, the method is applied to study the scaling behavior of epoxy-metal butt joints.

## 1. Introduction

Bi-material composites and structures are prevalent throughout a broad range of natural and engineered systems due to the ability of such systems to take advantage of the positive attributes of the individual constituents and to minimize their weaknesses. However, the interface between the two materials often remains as a critical region, thus limiting the overall performance of the composite. For systems that can be idealized as remaining essentially in the linear elastic regime, the stress intensity factors at the tips of any interfacial cracks or flaws can be used as controlling parameters. In the present paper, we develop a new boundary element formulation to evaluate these stress intensity factors for the two-dimensional case. The problem is complicated by the fact that the stress field around the interfacial crack tip is singular and log-periodic. This, in turn, necessitates the calculation of a complex stress intensity factor in which the usual opening and shear modes are intrinsically coupled.

Significant contributions toward understanding the physical and mathematical bi-material crack problem include the work by Muskhelishvili (1953), Williams (1959), Sih and Rice (1964), England (1965), Erdogan (1965), Rice and Sih (1965), Bogy (1971), Comninou (1977), Hutchinson *et al.* (1987) and Rice (1988). In terms of computational methods, Lin and Mar (1976) developed the first finite element formulation to analyze a crack between dissimilar materials, while Barsoum (1974) introduced the concept of quarter-point elements that has been used successfully for a range of fracture mechanics problems. Early work on the development of the boundary element method for linear elastic fracture mechanics includes that by Cruse (1978) and Blandford *et al.* (1981). The bi-material crack problem has been addressed more recently within the context of a boundary element method by Lee and Choi (1988), Yuuki and Cho (1989), Raveendra and Banerjee (1991), Lee (1996), Lim et al. (2002) and Zhou et al. (2005).

In the next section, we begin by providing the governing equations, along with an overview of the basic characteristics of the response for a bi-material crack in elastic media. This is followed by a presentation of the existing quarter-point boundary element fracture mechanics formulations, including the work by Raveendra and Banerjee (1991) for the bi-material problem. This sets the stage for our current development of a boundary element approach that explicitly addresses the singular, log-periodic behavior of the solution near the crack tip. Here we first consider the formulation for an elastic material bonded to a rigid body with a planar interfacial crack and then subsequently examine the more general bi-material problem. Afterwards, the results of several numerical examples are examined in comparison with analytical solutions and previously published approaches and physical experiments. The paper then finishes with some general conclusions.

**2. Governing equations**

For an isotropic, linear elastic bi-material boundary value problem, the governing Navier equations of equilibrium in the absence of body forces can be written:

$$(\lambda^{[\alpha]} + \mu^{[\alpha]})u_{j,ij}^{[\alpha]} + \mu^{[\alpha]}u_{i,jj}^{[\alpha]} = 0 \quad \text{in } V^{[\alpha]}, \tag{1}$$

where $\mathbf{u}^{[\alpha]}$ is the displacement vector, $\lambda^{[\alpha]}$ and $\mu^{[\alpha]}$ are the Lame elastic constants, and $V^{[\alpha]}$ represents the volume occupied by material $\alpha$ with $\alpha = 1,2$ for the two bodies. In (1) and subsequently, the usual indicial notation is employed for Latin indices. Besides these equilibrium equations, boundary conditions must be specified to form a well-posed problem. Here, we let

$$\mathbf{u}^{[\alpha]} = \tilde{\mathbf{u}}^{[\alpha]} \quad \text{on } S_u^{[\alpha]}, \tag{2a}$$

$$\mathbf{t}^{[\alpha]} = \tilde{\mathbf{t}}^{[\alpha]} \quad \text{on } S_t^{[\alpha]}, \tag{2b}$$

as the Dirichlet and Neumann conditions, respectively, with $\mathbf{t}^{[\alpha]}$ representing the surface tractions. Meanwhile on the interface between the two bodies

$$\mathbf{u}^{[1]} = \mathbf{u}^{[2]} \quad \text{on } S_b, \tag{3a}$$

$$\mathbf{t}^{[1]} = -\mathbf{t}^{[2]} \quad \text{on } S_b, \tag{3b}$$

and also

$$\mathbf{t}^{[\alpha]} = \mathbf{0} \quad \text{on } S_c^{[\alpha]}. \tag{3c}$$

With this definition, when an interface crack does exist, the crack surfaces are included in $S_c^{[\alpha]}$, while the remaining bonded portion of the interface forms $S_b$. Furthermore, we know from a local elastic eigen-analysis (Hutchinson *et al.*, 1987; Rice, 1988), the stresses $\boldsymbol{\sigma}$ at the crack tip are singular. Assuming a cracked planar interfacial surface, as shown in Fig. 1, the stresses behave in the following manner:

$$\lim_{r \to 0} \frac{(\sigma_{22} + i\sigma_{12})_{\theta=0}}{r^{-1/2 + i\varepsilon}} = \frac{K}{\sqrt{2\pi}}, \tag{4}$$

where $K = K_1 + iK_2$ is the complex stress intensity factor and

$$\varepsilon = \frac{1}{2\pi} \ln(\hat{\kappa}), \tag{5}$$

with

$$\hat{\kappa} = \frac{\kappa^{[1]} \mu^{[2]} + \mu^{[1]}}{\kappa^{[2]} \mu^{[1]} + \mu^{[2]}}, \tag{6}$$

$$\kappa^{[\alpha]} = \begin{cases} 3 - 4\nu^{[\alpha]} & \text{for plane strain} \\ \frac{3 - \nu^{[\alpha]}}{1 + \nu^{[\alpha]}}, & \text{for plane stress} \end{cases} \tag{7}$$

and Poisson ratio $\nu^{[\alpha]} = \lambda^{[\alpha]} / [2(\lambda^{[\alpha]} + \mu^{[\alpha]})]$.

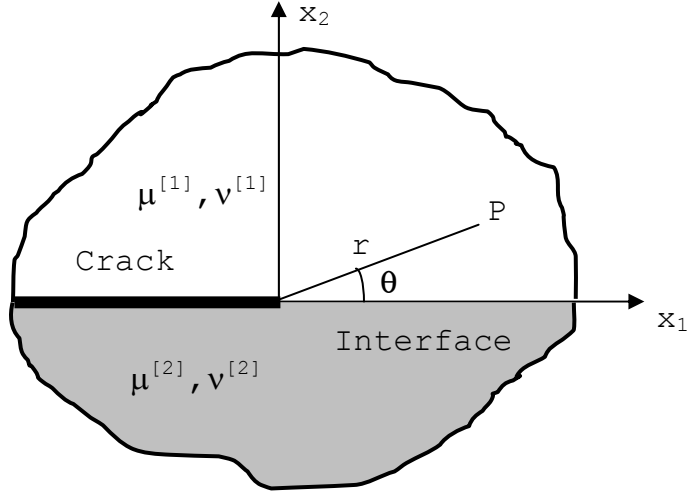

**Fig. 1.** Bi-material interfacial crack definition

Thus, for the general case, an oscillatory singular stress distribution obtains near the crack tip. Also, the convenient decomposition of response into opening (mode I) and shear (mode II) deformation patterns, which occurs in the cracked isotropic case, is no longer possible. Instead, the intrinsic coupling of these modes produces a characteristically distinct response for the bi-material case. An exception arises when the constitutive parameter $(1-2\nu)/\mu$ is identical for both materials. Then, the fracture modes again uncouple.

There are additional interesting features associated with a bi-material interface crack. For example, we should mention that the usual invariance of scale, which appears in the near field for the case of a crack in a single isotropic material, is replaced by discrete scale invariance for $\varepsilon \neq 0$ (Sornette, 1998). This suggests fundamentally different behavior of the bi-material interface crack and an association with many other critical phenomena exhibiting log-periodic character, such as percolation, biological growth and perhaps turbulence (Sornette, 1998). Much more insight is required to understand fully the physics of the problem. However, in the present work, our primary focus is on the accurate evaluation of the complex stress intensity factor *K* that characterizes the crack tip stress field. In the next section, we provide a review of some existing approaches.

## 3. Boundary integral formulations

In linear elastic fracture mechanics, boundary element methods eliminate the need for domain discretization and permit accurate resolution of surface-based field variables, thus providing an almost ideal computational setting for the evaluation of stress intensity factors (Banerjee, 1994; Watson, 1995; Cruse, 1996). The starting point for a boundary element formulation is an integral representation of the problem defined in (1)-(3). This can be written in the following form for each of the two material bodies:

$$C_{ij}^{[\alpha]}(\xi)u_j^{[\alpha]}(\xi) + \int_{S^{[\alpha]}} F_{ij}^{[\alpha]}(\xi,x)u_j^{[\alpha]}(x)\,dS(x) = \int_{S^{[\alpha]}} G_{ij}^{[\alpha]}(\xi,x)t_j^{[\alpha]}(x)\,dS(x), \tag{8}$$

where $G_{ij}^{[\alpha]}(\xi,x)$ and $F_{ij}^{[\alpha]}(\xi,x)$ are the singular Kelvin kernel functions, while $C_{ij}^{[\alpha]}(\xi)$ represents the local geometry at the point $\xi$. For plane strain,

$$G_{ij}(\xi,x) = \frac{1}{8\pi\mu(1-\nu)}\left[(3-4\nu)\delta_{ij}\ln\left(\frac{1}{r}\right) + z_i z_j\right] \tag{9a}$$

$$F_{ij}(\xi,x) = -\frac{1}{4\pi(1-\nu)r}\left[(1-2\nu)(z_i n_j - z_j n_i + \delta_{ij} z_k n_k) + 2z_i z_j z_k n_k\right] \tag{9b}$$

with $r^2 = (\xi_i - x_i)(\xi_i - x_i)$, $z_i = (\xi_i - x_i)/r$, and **n** representing the outward unit normal vector at each location $x$ on the boundary.

A traditional boundary element method (BEM) can be developed through the straightforward discretization of (8) for each body independently. For example, one may employ standard quadratic shape functions to represent both the boundary displacements and tractions, and then utilize the method of collocation to form a set of linear algebraic equations in the following form:

$$\mathbf{F}^{[\alpha]}\mathbf{U}^{[\alpha]} = \mathbf{G}^{[\alpha]}\mathbf{T}^{[\alpha]} \quad \text{for } \alpha = 1,2, \tag{10}$$

where now the $C_{ij}^{[\alpha]}$ coefficients are incorporated within the $\mathbf{F}^{[\alpha]}$ matrix. After collecting all of the unknowns in a vector **u** and applying the boundary and interface conditions, the discrete boundary element equations ultimately can be written in the following form:

$$\mathbf{Au} = \mathbf{v}. \tag{11}$$

and solved using either Gaussian elimination or a preconditioned iterative method.

However, for meaningful evaluation of the stress intensity factor, special treatment is required due to the singular nature of the near crack tip tractions. Blandford *et al.* (1981) introduced special quarter-point boundary elements in an attempt to resolve this difficulty. For the displacement field adjacent to the crack tip, quarter-point elements are employed using a formulation taken directly from the finite element literature (Barsoum, 1974). Thus, in two-dimensions, standard three-noded quadratic shape functions are utilized for the boundary displacement, but with the mid-nodes shifted to a position one-quarter of the element length from the crack tip. This provides the proper $\sqrt{r}$ displacement variation for cracks in a single isotropic material. On the other hand, because the traction variation for that same case is $1/\sqrt{r}$, Blandford et al. (1981) developed a traction-singular modified shape function to account for this variation within the element adjacent to the tip. Unfortunately, the resulting traction variables within the element, and particularly at the crack tip, usually do not provide reliable data for estimation of the stress intensity factor. Instead, crack-opening displacements are normally employed. Thus, for a single material crack oriented along the negative $x_1$-axis, the stress intensity factors are computed from the following relationship:

$$K_I + i K_{II} = \frac{\mu}{1+\kappa} \sqrt{\frac{2\pi}{r}} \left( \Delta u_2 + i \Delta u_1 \right), \tag{12}$$

where $K_I$ and $K_{II}$ are the mode I and mode II stress intensity factors, respectively, while $\Delta u_2$ and $\Delta u_1$ represent the corresponding crack opening displacements a distance $r$ from the tip.

The methodology outlined above is quite standard now, and a similar approach has been adopted for a range of fracture mechanics problems, including those involving thermoelastic behavior, dynamics and three-dimensional response.

One extension of this approach to bi-material interface cracks was examined by Raveendra and Banerjee (1991). In this work, the authors utilize the same quarter-point displacement and traction-singular elements to represent the surface variables adjacent to the crack tip. Thus, the characteristic oscillatory near-field behavior is not addressed explicitly in the element formulation. In fact, the boundary value problem is solved in exactly the same way as for a single material crack. However, the subsequent calculation of the stress intensity factor is modified to account for log-periodic response. The resulting post-processing operation can be written in the following form:

$$K = \frac{2(1+2i\varepsilon)\cosh(\pi\varepsilon)}{\beta^{[1]} + \beta^{[2]}} \exp(-i\varepsilon \ln r) \sqrt{\frac{2\pi}{r}} \left( \Delta u_2 + i \Delta u_1 \right), \tag{13}$$

where

$$\beta^{[\alpha]} = \frac{1+\kappa^{[\alpha]}}{\mu^{[\alpha]}}. \tag{14}$$

Notice that (13) reduces to (12) for the case with $\varepsilon = 0$.

## 4. Complex weighting function boundary integral formulations

*4.1. Introduction*

With this background in mind, we now develop a boundary integral formulation for the bi-material crack problem that explicitly considers the behavior of the deformation and stress fields near the tip. In particular, we account for the singular, log-periodic stress fields. However, before tackling the bi-material problem, we first examine the response of a single linear elastic isotropic medium bonded to a rigid body along a portion of the interface, since this also exhibits oscillatory singular behavior.

*4.2. Elastic-rigid interfacial crack*

Consider the idealized situation depicted in Fig. 2. From the corresponding elasticity solution, we find that the stresses near the crack tip again follow power law behavior with a complex power, as indicated in (4). The only exception occurs when the elastic medium is incompressible, in which case $\nu = 1/2$. Then from (5)-(7), we have $\varepsilon = 0$.

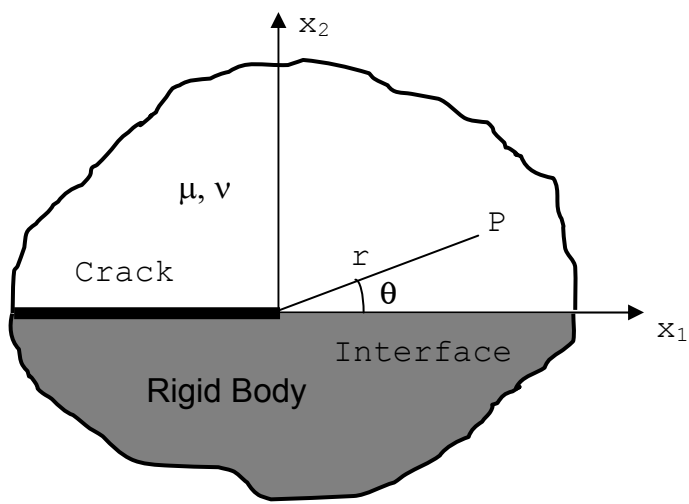

**Fig. 2.** Elastic-rigid interfacial crack definition

Using the Muskhelishvili-Kolosov formalism (Muskhelishvili, 1953), we may write the displacement and stress fields surrounding the crack in terms of complex potentials $\Phi$ and $\Psi$. That is,

$$2\mu(u_1 + iu_2) = \kappa\Phi(z) - z\overline{\Phi'(z)} - \overline{\Psi(z)} \tag{15}$$

$$\sigma_{11} + \sigma_{22} = 2\left(\Phi'(z) + \overline{\Phi'(z)}\right) \tag{16a}$$

$$\sigma_{22} - i\sigma_{12} = \Phi'(z) + \overline{\Phi'(z)} + z\overline{\Phi''(z)} + \overline{\Psi'(z)} \tag{16b}$$

where $z = re^{i\theta}$, primes denote derivatives with respect to $z$, and the overbars represent the complex conjugate. The potentials $\Phi$ and $\Psi$ are analytic everywhere, except at the crack tip. Therefore, we take

$$\Phi(z) = Az^{\lambda} + Bz^{\overline{\lambda}}, \tag{17a}$$

$$\Psi(z) = Cz^{\lambda} + Dz^{\overline{\lambda}}, \tag{17b}$$

and enforce the following homogeneous boundary conditions:

$$(u_1 + iu_2)_{\theta=0} = 0, \tag{18a}$$

$$(\sigma_{22} - i\sigma_{12})_{\theta=\pi} = 0. \tag{18b}$$

The corresponding characteristic equation becomes

$$e^{i2\pi\lambda} + \kappa = 0, \tag{19}$$

along with the following relationships

$$A = b + ic, \quad B = 0, \quad C = -\lambda A, \quad D = k\overline{A}, \tag{20}$$

where $b$ and $c$ are real-valued. Then, the eigenvalues can be written as

$$\lambda_n = n + \frac{1}{2} - i\varepsilon \quad \text{for} \quad n = 0, 1, 2, \ldots \tag{21}$$

with

$$\varepsilon = \frac{1}{2\pi}\ln(\kappa). \tag{22}$$

Notice that from (6) for $\mu^{[2]} \to \infty$, we have $\hat{\kappa} = \kappa^{[1]} = \kappa$. Thus, (22) is entirely consistent with the relation (5) for general bi-material interface cracks.

Of course, we are not only interested in the eigenvalues, but also the eigenfunctions. Thus, on the fixed side, we find

$$(u_1 + iu_2)_{\theta=0} = 0, \tag{23a}$$

$$(\sigma_{22} - i\sigma_{12})_{\theta=0} = \lambda A(1+\kappa)r^{\lambda-1} = -(t_2 - it_1)_{\theta=0}, \tag{23b}$$

while on the free side

$$2\mu(u_1 + iu_2)_{\theta=\pi} = \kappa A r^{\lambda}(e^{i\lambda\pi} - e^{-i\lambda\pi}) + \overline{\lambda}\overline{A}r^{\overline{\lambda}}\left[e^{-i\lambda\pi} - e^{-i(\overline{\lambda}-2)\pi}\right], \tag{24a}$$

$$-(t_2 - it_1)_{\theta=\pi} = 0. \tag{24b}$$

Next we define analytic weighted displacements **v** and weighted tractions **s** consistent with these eigenfunctions. More specifically, we let

$$2\mu \begin{Bmatrix} u_1 \\ u_2 \end{Bmatrix}\bigg|_{\theta=\pi} = \sqrt{\kappa}(1+\kappa)r^{1/2}\begin{bmatrix} -\cos(\varepsilon \ln r) & \sin(\varepsilon \ln r) \\ \sin(\varepsilon \ln r) & \cos(\varepsilon \ln r) \end{bmatrix}\begin{Bmatrix} v_1 \\ v_2 \end{Bmatrix} \quad (25a)$$

$$\begin{Bmatrix} t_1 \\ t_2 \end{Bmatrix}\bigg|_{\theta=0} = (1+\kappa)r^{-1/2}\left\{\begin{bmatrix} \frac{1}{2} & -\varepsilon \\ -\varepsilon & -\frac{1}{2} \end{bmatrix}\cos(\varepsilon \ln r) + \begin{bmatrix} -\varepsilon & -\frac{1}{2} \\ -\frac{1}{2} & \varepsilon \end{bmatrix}\sin(\varepsilon \ln r)\right\}\begin{Bmatrix} s_1 \\ s_2 \end{Bmatrix} \quad (25b)$$

where

$$\begin{Bmatrix} v_1 \\ v_2 \end{Bmatrix} = \sum_{n=0}^{\infty} (-1)^n r^n \begin{Bmatrix} c_n \\ b_n \end{Bmatrix} = \begin{Bmatrix} c_0 \\ b_0 \end{Bmatrix} - r\begin{Bmatrix} c_1 \\ b_1 \end{Bmatrix} + r^2\begin{Bmatrix} c_2 \\ b_2 \end{Bmatrix} - \cdots \quad (26a)$$

$$\begin{Bmatrix} s_1 \\ s_2 \end{Bmatrix} = \sum_{n=0}^{\infty} \frac{r^n}{\frac{1}{4}+\varepsilon^2}\begin{bmatrix} \frac{1}{2}(n+\frac{1}{2})+\varepsilon^2 & n\varepsilon \\ -n\varepsilon & \frac{1}{2}(n+\frac{1}{2})+\varepsilon^2 \end{bmatrix}\begin{Bmatrix} c_n \\ b_n \end{Bmatrix} \quad (26b)$$

Note that from (25) and (26), at $r=0$, we have $v_1(0)=s_1(0)$ and $v_2(0)=s_2(0)$. Furthermore, from (4), (23b) and (25b), we can write the complex stress intensity factor as

$$K = (2\pi)^{1/2}(1+\kappa)\left(\frac{1}{2}+i\varepsilon\right)(s_2(0)-is_1(0)) \quad (27)$$

Now we introduce these weighted variables into the boundary integral representation through a change of variable. Thus, for the single region with $\alpha=1$, (8) is rewritten as

$$C_{ij}(\xi)u_j(\xi) + \int_S F_{ij}(\xi,x)W_{jk}^u(x)v_k(x)dS(x) = \int_S G_{ij}(\xi,x)W_{jk}^t(x)s_k(x)dS(x) \quad (28)$$

where $\mathbf{W}^u$ and $\mathbf{W}^t$ are the weight matrices defined in the following manner:

$$\mathbf{W}^u = \frac{\sqrt{\kappa}(1+\kappa)}{2\mu}r^{1/2}\begin{bmatrix} -\cos(\varepsilon \ln r) & \sin(\varepsilon \ln r) \\ \sin(\varepsilon \ln r) & \cos(\varepsilon \ln r) \end{bmatrix} \quad (29a)$$

$$\mathbf{W}^t = (1+\kappa)r^{-1/2}\begin{bmatrix} \frac{1}{2} & -\varepsilon \\ -\varepsilon & -\frac{1}{2} \end{bmatrix}\cos(\varepsilon \ln r) + (1+\kappa)r^{-1/2}\begin{bmatrix} -\varepsilon & -\frac{1}{2} \\ -\frac{1}{2} & \varepsilon \end{bmatrix}\sin(\varepsilon \ln r) \quad (29b)$$

For the numerical implementation, we employ the usual quadratic shape functions to represent **v** and **s**. Thus,

$$\mathbf{v}(r) = N^1(r)\mathbf{V}^1 + N^2(r)\mathbf{V}^2 + N^3(r)\mathbf{V}^3 \tag{30a}$$

$$\mathbf{s}(r) = N^1(r)\mathbf{S}^1 + N^2(r)\mathbf{S}^2 + N^3(r)\mathbf{S}^3 \tag{30b}$$

with

$$N^1(r) = 2\eta^2 - 3\eta + 1, \tag{31a}$$

$$N^2(r) = 4\eta - 4\eta^2, \tag{31b}$$

$$N^3(r) = 2\eta^2 - \eta, \tag{31c}$$

and $\eta = r/L$ for an element of length $L$ adjacent to the crack tip.

*4.3. Bi-material interfacial crack*

Now we describe the extension of this methodology to the bi-material problem, illustrated initially in Fig. 1. Within the Muskhelishvili-Kolosov formalism, we define complex potentials $\Phi^{[\alpha]}$ and $\Psi^{[\alpha]}$, which are analytic everywhere in their corresponding domains, except at the crack tip. Therefore, we take for *Domain 1*

$$\Phi^{[1]}(z) = A_1 z^\lambda + B_1 z^{\overline{\lambda}}, \tag{32a}$$

$$\Psi^{[1]}(z) = C_1 z^\lambda + D_1 z^{\overline{\lambda}}, \tag{32b}$$

and for *Domain 2*

$$\Phi^{[2]}(z) = A_2 z^\lambda + B_2 z^{\overline{\lambda}}, \tag{32c}$$

$$\Psi^{[2]}(z) = C_2 z^\lambda + D_2 z^{\overline{\lambda}}, \tag{32d}$$

Then, after enforcing the following interface relationships and homogeneous boundary conditions:

$$(u_1 + iu_2)^{[1]}_{\theta=0} = (u_1 + iu_2)^{[2]}_{\theta=0}, \tag{33a}$$

$$(\sigma_{22} - i\sigma_{12})^{[1]}_{\theta=0} = (\sigma_{22} - i\sigma_{12})^{[2]}_{\theta=0} \tag{33b}$$

$$(\sigma_{22} - i\sigma_{12})^{[1]}_{\theta=\pi} = 0 \tag{33c}$$

$$(\sigma_{22} - i\sigma_{12})^{[2]}_{\theta=-\pi} = 0. \tag{33d}$$

we see that the terms involving the coefficients $B_1$ and $B_2$ could have been ignored from beginning, since $B_1 = B_2 = 0$. However, we should mention that these coefficients are not zero for the case of a general notch.

Then, the problem reduces to the solution of the characteristic equation

$$e^{i4\pi\lambda} - (1-\hat{\kappa})e^{i2\pi\lambda} - \hat{\kappa} = 0 \tag{34}$$

for $\hat{\kappa}$ defined as in (6), along with the relationships

$$\mu^{[2]}\left(\kappa^{[1]}A_1 - \bar{D}_1\right) = \mu^{[1]}\left(\kappa^{[2]}A_2 - \bar{D}_2\right) \tag{35a}$$

$$\mu^{[2]}(\lambda A_1 + C_1) = \mu^{[1]}(\lambda A_2 + C_2) \tag{35b}$$

$$A_1 + \bar{D}_1 = A_2 + \bar{D}_2 \tag{35c}$$

$$C_1 = -\lambda A_1 \tag{35d}$$

$$C_2 = -\lambda A_2 \tag{35e}$$

$$\bar{D}_1 = -A_1 e^{i2\pi\lambda} \tag{35f}$$

$$\bar{D}_2 = -A_2 e^{-i2\pi\lambda} \tag{35g}$$

The characteristic equation (34) decomposes to two equations:

$$e^{i2\pi\lambda} = 1 \tag{36a}$$

$$e^{i2\pi\lambda} = -\hat{\kappa} \tag{36b}$$

with corresponding sets of eigenvalues

$$\lambda_n = n \quad \text{for} \quad n = 0,1,2,\ldots \tag{37a}$$

$$\lambda_n = n + \frac{1}{2} - i\varepsilon \quad \text{for} \quad n = 0,1,2,\ldots \tag{37b}$$

where again

$$\varepsilon = \frac{1}{2\pi}\ln(\hat{\kappa}), \tag{38}$$

### 4.3.3   Set A: Integer eigenvalues and corresponding eigenmodes

We first consider modes associated with (36a) and (37a), which contain no singularity. In this case,

$$\frac{A_2}{A_1} = \frac{\mu^{[2]}\left(1+\kappa^{[1]}\right)}{\mu^{[1]}\left(1+\kappa^{[2]}\right)} = \frac{\beta^{[1]}}{\beta^{[2]}} \tag{39a}$$

$$C_1 = -\lambda A_1 \tag{39b}$$

$$C_2 = -\lambda A_2 \tag{39c}$$

$$\bar{D}_1 = -A_1 \tag{39d}$$

$$\bar{D}_2 = -A_2 \tag{39e}$$

In *Domain 1*, we have

$$2\mu^{[1]}(u_1 + iu_2)_n^{[1]} = (A_1)_n r^n \left[\kappa^{[1]} e^{in\theta} + e^{-in\theta}\right] + n(\bar{A}_1)_n r^n \left[e^{-in\theta} - e^{-i(n-2)\theta}\right] \tag{40a}$$

$$(\sigma_{11} + \sigma_{22})_n^{[1]} = 2n(A_1)_n r^{n-1} e^{i(n-1)\theta} + 2n(\bar{A}_1)_n r^{n-1} e^{-i(n-1)\theta} \tag{40b}$$

$$(\sigma_{22} - i\sigma_{12})_n^{[1]} = 2in(A_1)_n r^{n-1} \sin(n-1)\theta + n(n-1)(\bar{A}_1)_n r^{n-1} \left[e^{-i(n-3)\theta} - e^{-i(n-1)\theta}\right] \tag{40c}$$

Then, for the interface and free edges, we have

$$2\mu^{[1]}(u_1 + iu_2)_{n,\theta=0}^{[1]} = (A_1)_n (1+\kappa_1) r^n \tag{41a}$$

$$2\mu^{[1]}(u_1 + iu_2)_{n,\theta=\pi}^{[1]} = (-1)^n (A_1)_n (1+\kappa_1) r^n \tag{41b}$$

$$(\sigma_{11} + \sigma_{22})_{n,\theta=0}^{[1]} = 2n\left[(A_1)_n + (\bar{A}_1)_n\right] r^{n-1} \tag{41c}$$

$$(\sigma_{22} - i\sigma_{12})_{n,\theta=0}^{[1]} = 0 \tag{41d}$$

$$(\sigma_{11})_{n,\theta=0}^{[1]} = 2n\left[(A_1)_n + (\bar{A}_1)_n\right] r^{n-1} \tag{41e}$$

$$(\sigma_{11})_{n,\theta=\pi}^{[1]} = 2n\left[(A_1)_n + (\bar{A}_1)_n\right] (-r)^{n-1} \tag{41f}$$

Interestingly, from (41d), we find that there is no traction on the interface for this set of modes.

In *Domain 2*, we find

$$2\mu^{[2]}(u_1 + iu_2)_n^{[2]} = (A_2)_n r^n \left[\kappa_2 e^{in\theta} + e^{-in\theta}\right] + n(\bar{A}_2)_n r^n \left[e^{-in\theta} - e^{-i(n-2)\theta}\right] \tag{42a}$$

$$(\sigma_{11} + \sigma_{22})_n^{[2]} = 2n(A_2)_n r^{n-1} e^{i(n-1)\theta} + 2n(\bar{A}_2)_n r^{n-1} e^{-i(n-1)\theta} \tag{42b}$$

$$(\sigma_{22} - i\sigma_{12})_n^{[2]} = 2in(A_2)_n r^{n-1} \sin(n-1)\theta + n(n-1)(\bar{A}_2)_n r^{n-1} \left[e^{-i(n-3)\theta} - e^{-i(n-1)\theta}\right] \tag{42c}$$

and, consequently,

$$2\mu^{[2]}(u_1 + iu_2)_{n,\theta=0}^{[2]} = (A_2)_n (1+\kappa_2) r^n \tag{43a}$$

$$2\mu^{[2]}(u_1 + iu_2)_{n,\theta=\pi}^{[2]} = (-1)^n (A_2)_n (1+\kappa_2) r^n \tag{43b}$$

$$(\sigma_{22} - i\sigma_{12})_{n,\theta=0}^{[2]} = 0 \tag{43c}$$

From (41b) and (43b), we see that the displacements are the same on both free edges. Therefore, the responses of *Domains 1* and *2* are such that there is no discontinuity in the deformation on the interface and crack edges. Therefore, these modes represent a deformation of the body, as if there were no crack, which of course is compatible with having no singularity in these modes.

For the combination of these modes

$$(u_1 + iu_2)_{\theta=0} = \frac{1+\kappa^{[1]}}{2\mu^{[1]}} \sum_{n=0}^{\infty} (A_1)_n r^n \tag{44a}$$

$$(u_1 + iu_2)_{\theta=\pi,-\pi} = \frac{1+\kappa^{[1]}}{2\mu^{[1]}} \sum_{n=0}^{\infty} (A_1)_n (-1)^n r^n \tag{44b}$$

$$(\sigma_{22} - i\sigma_{12})_{\theta=0} = 0 \tag{44c}$$

which gives

$$\left\{ \begin{matrix} u_1 \\ u_2 \end{matrix} \right\}^A_{\theta=0} = \frac{1+\kappa^{[1]}}{2\mu^{[1]}} \left\{ \begin{matrix} b_n \\ c_n \end{matrix} \right\} r^n = \frac{1+\kappa^{[1]}}{2\mu^{[1]}} \left[ \left\{ \begin{matrix} b_0 \\ c_0 \end{matrix} \right\} + r \left\{ \begin{matrix} c_1 \\ b_1 \end{matrix} \right\} + r^2 \left\{ \begin{matrix} c_2 \\ b_2 \end{matrix} \right\} + \cdots \right] \tag{45a}$$

$$\left\{ \begin{matrix} u_1 \\ u_2 \end{matrix} \right\}^A_{\theta=\pi} = \frac{1+\kappa^{[1]}}{2\mu^{[1]}} \left[ \left\{ \begin{matrix} b_0 \\ c_0 \end{matrix} \right\} - r \left\{ \begin{matrix} c_1 \\ b_1 \end{matrix} \right\} + r^2 \left\{ \begin{matrix} c_2 \\ b_2 \end{matrix} \right\} - \cdots \right] \tag{45b}$$

$$\left\{ \begin{matrix} t_1 \\ t_2 \end{matrix} \right\}^A_{\theta=0} = 0 \tag{45c}$$

where the superscript $^A$ indicates modes from Set A. We find that the case $n=0$ corresponds to rigid body translations in the $x_1$ and $x_2$ directions, whereas the case $n=1$ is a combination of tension in the $x_1$ direction and rotation.

### 4.3.3   Set B: Complex eigenvalues and corresponding eigenmodes

Next, we examine modes associated with (36b) and (37b), which yield the relations

$$\frac{A_2}{A_1} = \frac{\mu^{[1]} + \mu^{[2]}\kappa^{[1]}}{\mu^{[2]} + \mu^{[1]}\kappa^{[2]}} = \hat{\kappa} \tag{46a}$$

$$C_1 = -\lambda A_1 \tag{46b}$$

$$C_2 = -\lambda A_2 \tag{46c}$$

$$\bar{D}_1 = -A_1 e^{i2\pi\lambda} \tag{46d}$$

$$\bar{D}_2 = -A_2 e^{-i2\pi\lambda} \tag{46e}$$

In *Domain 1*, we have

$$2\mu^{[1]}(u_1 + iu_2)_n^{[1]} = (A_1)_n r^{\lambda_n} \left[ \kappa_1 e^{i\lambda_n \theta} + e^{i2\pi\lambda_n} e^{-i\lambda_n \theta} \right] + \bar{\lambda}_n (\bar{A}_1)_n r^{\bar{\lambda}_n} \left[ e^{-i\bar{\lambda}_n \theta} - e^{-i(\bar{\lambda}_n - 2)\theta} \right] \tag{47a}$$

$$(\sigma_{22} - i\sigma_{12})_n^{[1]} = \lambda_n (A_1)_n r^{\lambda_n - 1} \left[ e^{i(\lambda_n - 1)\theta} + e^{i2\pi\lambda_n} e^{-i(\lambda_n - 1)\theta} \right]$$

$$+ \bar{\lambda}_n (\bar{\lambda}_n - 1)(\bar{A}_1)_n r^{\bar{\lambda}_n - 1} \left[ e^{-i(\bar{\lambda}_n - 3)\theta} - e^{-i(\bar{\lambda}_n - 1)\theta} \right] \tag{47b}$$

We see that the first eigenmode $n = 0$ in this set with eigenvalue

$$\lambda_0 = \frac{1}{2} - i\varepsilon \tag{48}$$

gives rise to the singularity of stress at the crack tip.

For general $n$, the interface and free edges respond as

$$(u_1 + iu_2)_{n,\theta=0}^{[1]} = (A_1)_n \frac{\kappa^{[1]}\kappa^{[2]} - 1}{2(\mu^{[2]} + \kappa^{[2]}\mu^{[1]})} r^{n+\frac{1}{2}-i\varepsilon} \tag{49a}$$

$$(u_1 + iu_2)_{n,\theta=\pi}^{[1]} = i(A_1)_n \frac{1 + \kappa^{[1]}}{2\mu^{[1]}} \sqrt{\hat{\kappa}}\, r^{n+\frac{1}{2}-i\varepsilon} \tag{49b}$$

$$(\sigma_{22} - i\sigma_{12})_{n,\theta=0}^{[1]} = (A_1)_n \left(n + \frac{1}{2} - i\varepsilon\right)(1 + \hat{\kappa}) r^{n-\frac{1}{2}-i\varepsilon} \tag{49c}$$

Therefore, for the overall combination of these modes, we find

$$(u_1 + iu_2)_{\theta=0}^{[1]} = \frac{\kappa^{[1]}\kappa^{[2]} - 1}{2(\mu^{[2]} + \kappa^{[2]}\mu^{[1]})} \sum_{n=0}^{\infty} (A_1)_n r^{n+\frac{1}{2}-i\varepsilon} \tag{50a}$$

$$(u_1 + iu_2)_{\theta=\pi}^{[1]} = i \frac{1 + \kappa^{[1]}}{2\mu^{[1]}} \sqrt{\hat{\kappa}} \sum_{n=0}^{\infty} (A_1)_n r^{n+\frac{1}{2}-i\varepsilon} \tag{50b}$$

$$(\sigma_{22} - i\sigma_{12})_{\theta=0}^{[1]} = (1 + \hat{\kappa}) \sum_{n=0}^{\infty} (A_1)_n \left(n + \frac{1}{2} - i\varepsilon\right) r^{n-\frac{1}{2}-i\varepsilon} \tag{50c}$$

In *Domain 2*, we obtain in general

$$2\mu^{[2]}(u_1+iu_2)_n^{[2]} = (A_2)_n r^{\lambda_n}\left[\kappa^{[2]}e^{i\lambda_n\theta}+e^{-i2\pi\lambda_n}e^{-i\lambda_n\theta}\right]$$
$$+\bar{\lambda}_n(\bar{A}_2)_n r^{\bar{\lambda}_n}\left[e^{-i\bar{\lambda}_n\theta}-e^{-i(\bar{\lambda}_n-2)\theta}\right] \tag{51a}$$

$$(\sigma_{22}-i\sigma_{12})_n^{[2]} = \lambda_n(A_2)_n r^{\lambda_n-1}\left[e^{i(\lambda_n-1)\theta}+e^{i2\pi\lambda_n}e^{-i(\lambda_n-1)\theta}\right]$$
$$+\bar{\lambda}_n(\bar{\lambda}_n-1)(\bar{A}_2)_n r^{\bar{\lambda}_n-1}\left[e^{-i(\bar{\lambda}_n-3)\theta}-e^{-i(\bar{\lambda}_n-1)\theta}\right] \tag{51b}$$

and for the interface and free edges, we have

$$(u_1+iu_2)_{n,\theta=0}^{[2]} = (A_2)_n \frac{e^{-i2\pi\lambda_n}+\kappa^{[2]}}{2\mu^{[2]}} r^{n+\frac{1}{2}-i\varepsilon} \tag{52a}$$

$$(u_1+iu_2)_{n,\theta=-\pi}^{[2]} = -i(A_2)_n \frac{1+\kappa^{[2]}}{2\mu^{[2]}}\sqrt{\hat{\kappa}}\, r^{n+\frac{1}{2}-i\varepsilon} \tag{52b}$$

which in terms of $(A_1)_n$ become

$$(u_1+iu_2)_{n,\theta=0}^{[2]} = (A_1)_n \frac{\kappa^{[1]}\kappa^{[2]}-1}{2\left(\mu^{[2]}+\kappa^{[2]}\mu^{[1]}\right)} r^{n+\frac{1}{2}-i\varepsilon} \tag{53a}$$

$$(u_1+iu_2)_{n,\theta=-\pi}^{[2]} = -i(A_1)_n \frac{1+\kappa^{[2]}}{2\mu^{[2]}}\sqrt{\hat{\kappa}}\, r^{n+\frac{1}{2}-i\varepsilon} \tag{53b}$$

Therefore, for the combination of these modes, we have

$$(u_1+iu_2)_{\theta=0}^{[1,2]} = \frac{\kappa^{[1]}\kappa^{[2]}-1}{2\left(\mu^{[2]}+\kappa^{[2]}\mu^{[1]}\right)} r^{\frac{1}{2}-i\varepsilon} \sum_{n=0}^{\infty}(A_1)_n r^n \tag{54a}$$

$$(u_1+iu_2)_{\theta=\pi}^{[1]} = \frac{1+\kappa^{[1]}}{2\mu^{[1]}}\sqrt{\hat{\kappa}}\, r^{\frac{1}{2}-i\varepsilon}\sum_{n=0}^{\infty}i(A_1)_n r^n \tag{54b}$$

$$(u_1+iu_2)_{\theta=-\pi}^{[2]} = -\frac{1+\kappa^{[2]}}{2\mu^{[2]}}\sqrt{\hat{\kappa}}\, r^{\frac{1}{2}-i\varepsilon}\sum_{n=0}^{\infty}i(A_1)_n r^n \tag{54c}$$

$$(\sigma_{22}-i\sigma_{12})_{\theta=0}^{[1,2]} = (1+\hat{\kappa})r^{-\frac{1}{2}-i\varepsilon}\left(\frac{1}{2}-i\varepsilon\right)\sum_{n=0}^{\infty}\frac{\frac{1}{2}\left(n+\frac{1}{2}\right)+\varepsilon^2+in\varepsilon}{\frac{1}{4}+\varepsilon^2}(A_1)_n r^n \tag{54d}$$

or, in vector form, the interfacial displacements can be written

$$\left\{ \begin{array}{l} u_1 \\ u_2 \end{array} \right\}^{[1,2]}_{\theta=0} = \frac{\kappa^{[1]}\kappa^{[2]}-1}{2\left(\mu^{[2]}+\kappa^{[2]}\mu^{[1]}\right)} r^{1/2} \begin{bmatrix} \sin(\varepsilon\ln r) & \cos(\varepsilon\ln r) \\ \cos(\varepsilon\ln r) & -\sin(\varepsilon\ln r) \end{bmatrix} \sum_{n=0}^{\infty} r^n \left\{ \begin{array}{l} c_n \\ b_n \end{array} \right\} \qquad (55a)$$

$$\left\{ \begin{array}{l} u_1 \\ u_2 \end{array} \right\}^{[1]}_{\theta=\pi} = \frac{1+\kappa^{[1]}}{2\mu^{[1]}} \sqrt{\hat{\kappa}}\, r^{1/2} \begin{bmatrix} -\cos(\varepsilon\ln r) & \sin(\varepsilon\ln r) \\ \sin(\varepsilon\ln r) & \cos(\varepsilon\ln r) \end{bmatrix} \sum_{n=0}^{\infty} r^n \left\{ \begin{array}{l} c_n \\ b_n \end{array} \right\} \qquad (55b)$$

$$\left\{ \begin{array}{l} u_1 \\ u_2 \end{array} \right\}^{[2]}_{\theta=-\pi} = \frac{1+\kappa^{[2]}}{2\mu^{[2]}} \sqrt{\hat{\kappa}}\, r^{1/2} \begin{bmatrix} \cos(\varepsilon\ln r) & -\sin(\varepsilon\ln r) \\ -\sin(\varepsilon\ln r) & -\cos(\varepsilon\ln r) \end{bmatrix} \sum_{n=0}^{\infty} r^n \left\{ \begin{array}{l} c_n \\ b_n \end{array} \right\} \qquad (55c)$$

and the traction on the interface acting on *Domain 2* is

$$\left\{ \begin{array}{l} t_1 \\ t_2 \end{array} \right\}^{[2]}_{\theta=0} = \left\{ \begin{array}{l} \sigma_{12} \\ \sigma_{22} \end{array} \right\}_{\theta=0} = (1+\hat{\kappa}) r^{-1/2} \begin{bmatrix} -\frac{1}{2}\cos(\varepsilon\ln r)+\varepsilon\sin(\varepsilon\ln r) & \varepsilon\cos(\varepsilon\ln r)+\frac{1}{2}\sin(\varepsilon\ln r) \\ \varepsilon\cos(\varepsilon\ln r)+\frac{1}{2}\sin(\varepsilon\ln r) & \frac{1}{2}\cos(\varepsilon\ln r)-\varepsilon\sin(\varepsilon\ln r) \end{bmatrix}$$

$$\times \sum_{n=0}^{\infty} \frac{r^n}{\frac{1}{4}+\varepsilon^2} \begin{bmatrix} \frac{1}{2}(n+\frac{1}{2})+\varepsilon^2 & n\varepsilon \\ -n\varepsilon & \frac{1}{2}(n+\frac{1}{2})+\varepsilon^2 \end{bmatrix} \left\{ \begin{array}{l} c_n \\ b_n \end{array} \right\} \qquad (55d)$$

Finally, we obtain

$$\left\{ \begin{array}{l} u_1 \\ u_2 \end{array} \right\}^{[1,2],B}_{\theta=0} = \frac{\kappa^{[1]}\kappa^{[2]}-1}{2\left(\mu^{[2]}+\kappa^{[2]}\mu^{[1]}\right)} r^{1/2} \left\{ \begin{bmatrix} 0 & 1 \\ 1 & 0 \end{bmatrix} \cos(\varepsilon\ln r) + \begin{bmatrix} 1 & 0 \\ 0 & -1 \end{bmatrix} \sin(\varepsilon\ln r) \right\} \left\{ \begin{array}{l} v_1 \\ v_2 \end{array} \right\} \qquad (56a)$$

$$\left\{ \begin{array}{l} u_1 \\ u_2 \end{array} \right\}^{[1],B}_{\theta=\pi} = \frac{1+\kappa^{[1]}}{2\mu^{[1]}} \sqrt{\hat{\kappa}}\, r^{1/2} \left\{ \begin{bmatrix} -1 & 0 \\ 0 & 1 \end{bmatrix} \cos(\varepsilon\ln r) + \begin{bmatrix} 0 & 1 \\ 1 & 0 \end{bmatrix} \sin(\varepsilon\ln r) \right\} \left\{ \begin{array}{l} v_1 \\ v_2 \end{array} \right\} \qquad (56b)$$

$$\left\{ \begin{array}{l} u_1 \\ u_2 \end{array} \right\}^{[2],B}_{\theta=-\pi} = \frac{1+\kappa^{[2]}}{2\mu^{[2]}} \sqrt{\hat{\kappa}}\, r^{1/2} \left\{ \begin{bmatrix} 1 & 0 \\ 0 & -1 \end{bmatrix} \cos(\varepsilon\ln r) + \begin{bmatrix} 0 & -1 \\ -1 & 0 \end{bmatrix} \sin(\varepsilon\ln r) \right\} \left\{ \begin{array}{l} v_1 \\ v_2 \end{array} \right\} \qquad (56c)$$

$$\left\{ \begin{array}{l} t_1 \\ t_2 \end{array} \right\}^{[1],B}_{\theta=0} = -\left\{ \begin{array}{l} t_1 \\ t_2 \end{array} \right\}^{[2],B}_{\theta=0} = (1+\hat{\kappa}) r^{-1/2} \left\{ \begin{bmatrix} \frac{1}{2} & -\varepsilon \\ -\varepsilon & -\frac{1}{2} \end{bmatrix} \cos(\varepsilon\ln r) + \begin{bmatrix} -\varepsilon & -\frac{1}{2} \\ -\frac{1}{2} & \varepsilon \end{bmatrix} \sin(\varepsilon\ln r) \right\} \left\{ \begin{array}{l} s_1 \\ s_2 \end{array} \right\} \qquad (56d)$$

where the superscript $^B$ indicates modes from Set B and

$$\left\{ \begin{array}{l} v_1 \\ v_2 \end{array} \right\} = \sum_{n=0}^{\infty} r^n \left\{ \begin{array}{l} c_n \\ b_n \end{array} \right\} = \left\{ \begin{array}{l} c_0 \\ b_0 \end{array} \right\} + r \left\{ \begin{array}{l} c_1 \\ b_1 \end{array} \right\} + r^2 \left\{ \begin{array}{l} c_2 \\ b_2 \end{array} \right\} + \cdots \qquad (57a)$$

$$\left\{ \begin{array}{c} s_1 \\ s_2 \end{array} \right\} = \sum_{n=0}^{\infty} \frac{r^n}{\frac{1}{4}+\varepsilon^2} \begin{bmatrix} \frac{1}{2}(n+\frac{1}{2})+\varepsilon^2 & n\varepsilon \\ -n\varepsilon & \frac{1}{2}(n+\frac{1}{2})+\varepsilon^2 \end{bmatrix} \left\{ \begin{array}{c} c_n \\ b_n \end{array} \right\} \tag{57b}$$

Notice again that at the crack tip, $v_1(0) = s_1(0)$ and $v_2(0) = s_2(0)$. Meanwhile, the complex stress intensity factor can be written as

$$K = (2\pi)^{1/2}(1+\hat{\kappa})\left(\frac{1}{2}+i\epsilon\right)(s_2(0) - is_1(0)) \tag{58}$$

### 4.3.3 Overall response

The combination of both sets of eigenmodes gives the general elastic solution as

$$u_j = u_j^A + u_j^B \tag{59a}$$

$$t_j = t_j^A + t_j^B \tag{59b}$$

As was the case for the elastic-rigid interfacial crack of Section 4.2, weighted displacements and tractions become the primary variables on the interface. However, we must also include a linear displacement term, where $u_j^A = U_j^A + a_{jk}x_k$ in which $U_j^A$ is crack tip displacement and $a_{jk}x_k$ is a combination of tension in the $x_1$ direction and rotation. Therefore,

$$u_j = U_j^A + a_{jk}x_k + W_{jk}^u(x)v_k \tag{60a}$$

$$t_j = W_{jk}^t(x)s_k \tag{60b}$$

where these weights are defined such that at the crack tip

$$v_k(0) = s_k(0) \tag{61}$$

Thus, the weight matrices $\mathbf{W}^u$ and $\mathbf{W}^t$ are defined in the following manner:

$$W_{\theta=0}^u = \frac{\kappa^{[1]}\kappa^{[2]}-1}{2\left(\mu^{[2]}+\kappa^{[2]}\mu^{[1]}\right)}r^{1/2}\left\{\begin{bmatrix} 0 & 1 \\ 1 & 0 \end{bmatrix}\cos(\varepsilon \ln r) + \begin{bmatrix} 1 & 0 \\ 0 & -1 \end{bmatrix}\sin(\varepsilon \ln r)\right\} \tag{62a}$$

$$W_{\theta=\pi}^u = \frac{1+\kappa^{[1]}}{2\mu^{[1]}}\sqrt{\hat{\kappa}}\, r^{1/2}\left\{\begin{bmatrix} -1 & 0 \\ 0 & 1 \end{bmatrix}\cos(\varepsilon \ln r) + \begin{bmatrix} 0 & 1 \\ 1 & 0 \end{bmatrix}\sin(\varepsilon \ln r)\right\} \tag{62b}$$

$$W_{\theta=-\pi}^u = \frac{1+\kappa^{[2]}}{2\mu^{[2]}}\sqrt{\hat{\kappa}}\, r^{1/2}\left\{\begin{bmatrix} 1 & 0 \\ 0 & -1 \end{bmatrix}\cos(\varepsilon \ln r) + \begin{bmatrix} 0 & -1 \\ -1 & 0 \end{bmatrix}\sin(\varepsilon \ln r)\right\} \tag{62c}$$

$$W_{\theta=0}^t = (1+\hat{\kappa})r^{-1/2}\left\{\begin{bmatrix} \frac{1}{2} & -\varepsilon \\ -\varepsilon & -\frac{1}{2} \end{bmatrix}\cos(\varepsilon \ln r) + \begin{bmatrix} -\varepsilon & -\frac{1}{2} \\ -\frac{1}{2} & \varepsilon \end{bmatrix}\sin(\varepsilon \ln r)\right\} \tag{62d}$$

and the boundary integral representation (8) is rewritten for *Domain* $\alpha$ as

$$C_{ij}^{[\alpha]}(\xi)u_j^{[\alpha]}(\xi) + \int_{S^{[\alpha]}} F_{ij}^{[\alpha]}(\xi,x)W_{jk}^u(x)v_k^{[\alpha]}(x)dS(x) = \int_{S^{[\alpha]}} G_{ij}^{[\alpha]}(\xi,x)W_{jk}^t(x)s_k^{[\alpha]}(x)dS(x) \quad (63)$$

## 5. Non-standard Gaussian quadrature

In addition to the usual $\ln r$ and $1/r$ terms that appear in (8), the evaluation of integrals in (28) and (63) involve singular terms involving $r^\gamma \cos(\varepsilon \ln r)$, $r^\gamma \sin(\varepsilon \ln r)$, $r^\gamma \ln r \cos(\varepsilon \ln r)$ and $r^\gamma \ln r \sin(\varepsilon \ln r)$, where $\gamma = \pm 1/2$. Therefore, special Gaussian quadrature formulas with non-classical weights $W(x)$ are developed, such that

$$\int_0^1 W(x)f(x)dx = \sum_{j=1}^N w_j f(x_j) \quad (64)$$

The development of these *N*-point formulas requires determining sets of abscissas $x_j$ and weights $w_j$ and follows concepts presented in Press *et al.* (1992).

The idea is based on the set of orthogonal polynomials $p_j(x)$ with respect to the weight function $W(x)$ with recurrence relation

$$p_{-1}(x) = 0 \quad (65a)$$

$$p_0(x) = 1 \quad (65b)$$

$$p_{j+1}(x) \equiv (x - a_j)p_j(x) - b_j p_{j-1}(x) \quad j = 0, 1, 2, \cdots \quad (65c)$$

where

$$a_j = \frac{\langle xp_j | p_j \rangle}{\langle p_j | p_j \rangle} \quad j = 0, 1, \cdots \quad (66a)$$

$$b_j = \frac{\langle p_j | p_j \rangle}{\langle p_{j-1} | p_{j-1} \rangle} \quad j = 1, 2, \cdots \quad (66b)$$

The coefficient $b_0$ is arbitrary, and is taken here as zero. The symbol $\langle f | g \rangle$ represents the scalar product of functions $f(x)$ and $g(x)$ with respect to the weight function $W(x)$ over the interval $(0, 1)$, that is

$$\langle f | g \rangle = \int_0^1 W(x)f(x)g(x)dx \quad (67)$$

For Gaussian quadrature, the abscissas of the *N*-point formulas with weighting function $W(x)$ are the roots of the orthogonal polynomial $p_N(x)$. Once we know the abscissas $x_1, \cdots, x_N$, we can find the weights $w_j$, $j = 1, \cdots, N$. However, when the recurrence relations (65) are known, the best way to find these roots is using the algorithm of Golub and Welsch (1969), which shows the roots are the eigenvalues of the symmetric tridiagonal Jacobi matrix **J** defined by

$$\mathbf{J} = \begin{bmatrix} a_0 & \sqrt{b_1} & & & \\ \sqrt{b_1} & a_1 & \sqrt{b_2} & & \\ & \vdots & & \vdots & \\ & & \sqrt{b_{N-2}} & a_{N-2} & \sqrt{b_{N-1}} \\ & & & \sqrt{b_{N-1}} & a_{N-1} \end{bmatrix} \tag{68}$$

Then, the abscissas $x_j$ of the Gaussian quadratures are the eigenvalues $x$ of the eigenproblem

$$\mathbf{J}\mathbf{\psi} = x\mathbf{\psi} \tag{69}$$

where $\mathbf{\psi}_j$ is the eigenvector corresponding to the eigenvalue $x_j$. If this eigenvector is normalized so that $\mathbf{\psi} \bullet \mathbf{\psi} = 1$, the corresponding weights are

$$w_j = \mu_0 \psi_{j1}^2 \tag{70}$$

where

$$\mu_0 = \int_0^1 W(x) dx \tag{71}$$

and $\psi_{j1}$ is the first component of $\mathbf{\psi}_j$.

For a non-standard weight function, the orthogonal polynomial $p_j(x)$ and the coefficients $a_j$ and $b_j$ of the recurrence relations (65) are not known beforehand. An efficient way to find these is by using Sack and Donovan (1972) method, which uses known orthogonal functions $\pi_j(x)$ with known recurrence relations analogous to (65) as

$$\pi_{-1}(x) = 0 \tag{72a}$$

$$\pi_0(x) = 1 \tag{72b}$$

$$\pi_{j+1}(x) \equiv (x - \alpha_j)\pi_j(x) - \beta_j \pi_{j-1}(x) \quad j = 0, 1, 2, \cdots \tag{72c}$$

Here the coefficients $\alpha_j$ and $\beta_j$ are known explicitly, along with the modified moments

$$v_j = \int_0^1 \pi_j(x) W(x) dx \quad j = 0, 1, \cdots, 2N-1 \tag{73}$$

An efficient algorithm by Wheeler (1974) can be used to find $a_j$ and $b_j$ is via a set of intermediate quantities

$$\sigma_{k,l} = \langle p_k | \pi_l \rangle \quad k, l \geq -1 \tag{74}$$

Press *et al.* (1992) define the algorithm as follows. First, initialize

$$\sigma_{-1,l} = 0 \quad l = 1, 2, \cdots, 2N-2 \tag{75a}$$

$$\sigma_{0,l} = v_l \quad l = 0, 1, \cdots, 2N-1 \tag{75b}$$

$$a_0 = \alpha_0 + \frac{v_1}{v_0} \tag{75c}$$

$$b_0 = 0 \tag{75d}$$

Then, for $k = 1, 2, \cdots, N-1$, we compute

$$\sigma_{k,l} = \sigma_{k-1,l+1} - (a_{k-1} - \alpha_l)\sigma_{k-1,l} - b_{k-1}\sigma_{k-2,l} + \beta_l \sigma_{k-1,l-1} \quad l = k, k+1, \cdots, 2N-k-1 \tag{76}$$

Finally, we obtain

$$a_k = \alpha_k - \frac{\sigma_{k-1,k}}{\sigma_{k-1,k-1}} + \frac{\sigma_{k,k+1}}{\sigma_{k,k}} \tag{77a}$$

$$b_k = \frac{\sigma_{k,k}}{\sigma_{k-1,k-1}} \tag{77b}$$

Here, for auxillary orthogonal polynomials $\pi_j(x)$, we choose the shifted Legendre polynomials in monic form

$$\pi_j(x) = \frac{(j!)^2}{(2j)!} P_j(2x-1) \tag{78}$$

where the

$$P_j(2x-1) = (-1)^j \sum_{k=0}^{j} \frac{(j+k)!}{(k!)^2 (j-k)!} (-x)^k \tag{79}$$

are orthogonal in the interval $(0, 1)$. Therefore,

$$\pi_j = \frac{(j!)^2}{(2j)!} (-1)^j \sum_{k=0}^{j} (-1)^k \frac{(j+k)!}{(k!)^2 (j-k)!} x^k \tag{80}$$

For this case, the coefficients $\alpha_j$ and $\beta_j$ in the recurrence relation (72) are

$$\alpha_j = \frac{1}{2} \quad j = 0, 1, \cdots \tag{81a}$$

$$\beta_j = \frac{1}{4(4-j^{-2})} \quad j = 1, 2, \cdots \tag{81b}$$

and $\beta_0$ is taken to be zero.

For the our present cases with weighting functions

$$W(x) = \begin{cases} x^\gamma \cos(\varepsilon \ln x) \\ x^\gamma \sin(\varepsilon \ln x) \end{cases} \tag{82}$$

the modified moments $v_j$ can be obtained by using the complex weight function

$$W(x) = x^{\gamma + i\varepsilon} \tag{83}$$

Therefore, the complex modified moment is

$$v_j = \int_0^1 \pi_j(x) x^{\gamma+i\varepsilon} dx \quad j = 0, 1, \cdots, 2N-1$$

$$v_j = \int_0^1 \frac{(j!)^2}{(2j)!}(-1)^j \sum_{k=0}^{j}(-1)^k \frac{(j+k)!}{(k!)^2(j-k)!} x^k x^{\gamma+i\varepsilon} dx$$

$$v_j = \frac{(j!)^2}{(2j)!}(-1)^j \sum_{k=0}^{j}(-1)^k \frac{(j+k)!}{(k!)^2(j-k)!} \int_0^1 x^{\gamma+k+i\varepsilon} dx$$

$$v_j = \frac{(j!)^2}{(2j)!}(-1)^j \sum_{k=0}^{j}(-1)^k \frac{(j+k)!}{(k!)^2(j-k)!} \frac{1}{\gamma+k+1+i\varepsilon} \tag{84}$$

and finally

$$v_j = \frac{(j!)^2}{(2j)!}(-1)^j \sum_{k=0}^{j}(-1)^k \frac{(j+k)!}{(k!)^2(j-k)!}\left[\frac{\gamma+k+1}{(\gamma+k+1)^2+\varepsilon^2} - i\frac{\varepsilon}{(\gamma+k+1)^2+\varepsilon^2}\right] \tag{85}$$

The real and imaginary parts of this complex expression are the required modified moments. Thus,

$$v_j = \begin{cases} \dfrac{(j!)^2}{(2j)!}(-1)^j \sum_{k=0}^{j}(-1)^k \dfrac{(j+k)!}{(k!)^2(j-k)!}\dfrac{\gamma+k+1}{(\gamma+k+1)^2+\varepsilon^2} \\[2ex] -\dfrac{(j!)^2}{(2j)!}(-1)^j \sum_{k=0}^{j}(-1)^k \dfrac{(j+k)!}{(k!)^2(j-k)!}\dfrac{\varepsilon}{(\gamma+k+1)^2+\varepsilon^2} \end{cases} \tag{86}$$

Similarly, the modified moments for the cases

$$W(x) = \begin{cases} x^\gamma \ln x \cos(\varepsilon \ln x) \\ x^\gamma \ln x \sin(\varepsilon \ln x) \end{cases} \tag{87}$$

are obtained by using the complex weight function

$$W(x) = x^{\gamma+i\varepsilon} \ln x \tag{88}$$

Therefore, the corresponding complex modified moment can be written

$$v_j = \int_0^1 \frac{(j!)^2}{(2j)!}(-1)^j \sum_{k=0}^{j}(-1)^k \frac{(j+k)!}{(k!)^2(j-k)!} x^k x^{\gamma+i\varepsilon} \ln x\, dx$$

$$v_j = \frac{(j!)^2}{(2j)!}(-1)^j \sum_{k=0}^{j}(-1)^k \frac{(j+k)!}{(k!)^2(j-k)!}\left[-\frac{1}{(\gamma+k+1+i\varepsilon)^2}\right]$$

$$v_j = \frac{(j!)^2}{(2j)!}(-1)^j \sum_{k=0}^{j}(-1)^k \frac{(j+k)!}{(k!)^2(j-k)!}\left[-\frac{(\gamma+k+1)^2-\varepsilon^2}{\left[(\gamma+k+1)^2+\varepsilon^2\right]^2}+i\frac{2(\gamma+k+1)\varepsilon}{(\gamma+k+1)^2+\varepsilon^2}\right] \tag{89}$$

The real and imaginary parts of this complex expression are the desired modified moments. Thus,

$$v_j = \begin{cases} -\dfrac{(j!)^2}{(2j)!}(-1)^j \sum_{k=0}^{j}(-1)^k \dfrac{(j+k)!}{(k!)^2(j-k)!}\dfrac{(\gamma+k+1)^2-\varepsilon^2}{\left[(\gamma+k+1)^2+\varepsilon^2\right]^2} \\ \\ \dfrac{(j!)^2}{(2j)!}(-1)^j \sum_{k=0}^{j}(-1)^k \dfrac{(j+k)!}{(k!)^2(j-k)!}\dfrac{2(\gamma+k+1)\varepsilon}{(\gamma+k+1)^2+\varepsilon^2} \end{cases} \tag{90}$$

Knowing coefficients $\alpha_j$, $\beta_j$ and modified moments $v_j$ from (86) and (90), we can obtain the coefficients $a_j$ and $b_j$ via intermediate quantities $\sigma_{k,l}$. Then, by forming the corresponding Jacobi matrix **J** in (69), the set of abscissas $x_j$ and weights $w_j$ are determined by solving the corresponding eigenvalue problem. These non-standard Gaussian quadrature formulas are then used to evaluate the singular integrals appearing in (28) and (63). Additionally, in order to maintain accuracy, sub-segmentation must be used for elements containing kernel and crack tip singularities at multiple nodes.

## 6. Numerical examples

*6.1. Introduction*

In this section, several example problems will be presented to illustrate the performance of the proposed methodology. The examples include both single material and bi-material interface problems of linear elastic fracture mechanics. The final example involves analysis of an epoxy-steel butt joint to determine the scaling behavior. Solutions are then compared with results from the physical experiments by Reedy and Guess (1993, 1997).

*6.2. Problem 1: Circular elastic sector*

As an initial example, consider the elastic circular sector bonded to a rigid half-space along the positive $x_1$-axis, as defined in Fig. 3. Dirichlet boundary conditions, associated with the eigenmode for $n = 0$, are specified on the entire outer circumference at $r = a$, such that the analytical solution all along the bonded interface becomes $s_1 = 1$ and $s_2 = 1$. Consider first the case with $a = 1$, $\mu = 1$ and $\nu = 0.30$. Then, from (22), $\varepsilon = 0.0935492$ and the complex stress intensity factor from (27) becomes $K = K_1 + iK_2 = 4.16586 - 2.85270i$.

A total of eighty-six quadratic boundary elements are employed for the numerical analysis. The results provided in Fig. 4 show the variation of weighted traction components $s_1$ and $s_2$ along the interface and indicate that very accurate solutions are obtained with the present boundary element formulation. Detailed quantitative comparisons of weighted crack tip tractions and stress intensity factors are presented in Tables 1 and 2, respectively. The latter table includes results for $\nu = 0.30$ and for the incompressible case with $\nu = 0.50$. All of these data reveal that the present boundary element solutions are accurate to approximately four digits.

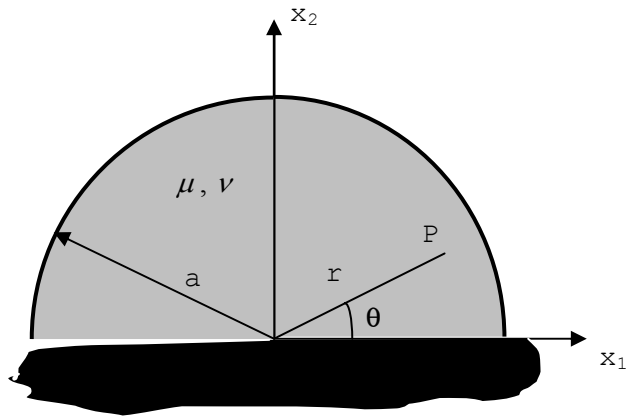

**Fig. 3.** Problem 1 Circular sector definition

**Table 1.** Problem 1 weighted tractions at tip ( $a=1, \mu=1, \nu=0.30$ )

|  | $s_1(0)$ | $s_2(0)$ |
|---|---|---|
| Exact | 1.0000 | 1.0000 |
| BEM (weighted tractions) | 0.99997 | 1.0000 |

**Table 2.** Problem 1 stress intensity factors ( $a=1, \mu=1$ )

| $\nu=0.30$ | $K_1$ | $K_2$ |
|---|---|---|
| Exact | 4.16586 | -2.85270 |
| BEM (weighted tractions) | 4.16584 | -2.85259 |
| BEM (crack opening displacements) | 4.16603 | -2.85250 |
| $\nu=0.50$ | $K_1$ | $K_2$ |
| Exact | 2.50663 | -2.50663 |
| BEM (weighted tractions) | 2.50688 | -2.50650 |
| BEM (crack opening displacements) | 2.50683 | -2.50658 |

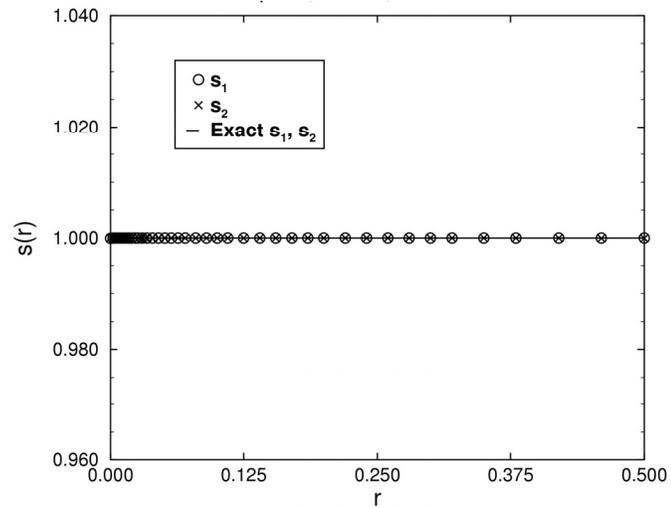

**Fig. 4.** Problem 1 circular sector results ( $a=1, \mu=1, \nu=0.30$ )

*6.3. Problem 2: Elastic square plate*

The second example problem involves a square plate perfectly bonded to a rigid half-space along a portion its lower edge. All problem parameters are provided in Fig. 5.

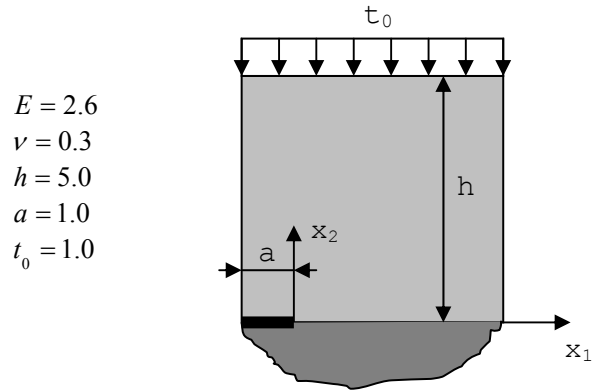

**Fig. 5.** Problem 2 square plate definition

Results of a convergence study are provided in Fig. 6. Here the weighted traction components $s_1$ and $s_2$ are plotted versus distance $r$ from the crack tip. Mesh A contains only eighteen quadratic boundary elements to model the entire plate, while each refinement to create Mesh B and then Mesh C involves a doubling of the number of elements. The overall convergence is evidently quite good. However, a more detailed view for $s_2(r)$ is presented in Fig. 7 for nodes near the crack tip. Two additional levels of refinement are also included, again with a doubling of the number of elements between the levels. A clear picture of the convergence is apparent in this latter figure.

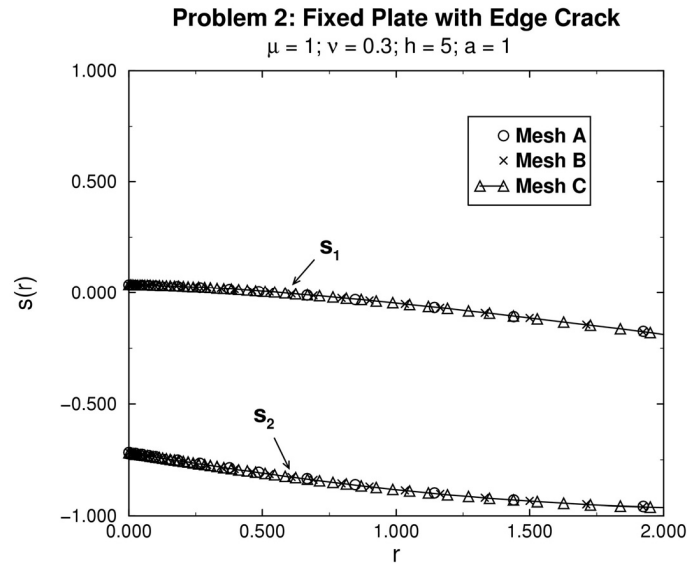

**Fig. 6.** Problem 2 weighted tractions convergence study

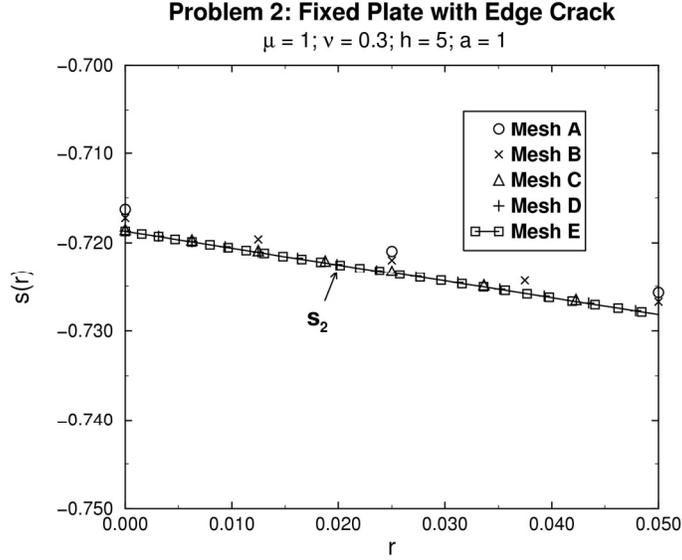

**Fig. 7.** Problem 2 weighted traction $s_2$ solutions near crack tip

*6.4. Problem 3: Bonded elastic circular sectors*

The third problem involves a cracked interface between two dissimilar elastic circular sectors, as illustrated in Fig. 8. For this problem, two cases were considered involving the specification of either Dirichlet or Neumann boundary conditions for the first singular eigenmode with $\lambda_0 = 1/2 - i\varepsilon$. Furthermore, we consider $(A_1)_0 = 1 + 1i$, such that

$$\begin{Bmatrix} s_1 \\ s_2 \end{Bmatrix} = \begin{Bmatrix} 1 \\ 1 \end{Bmatrix}$$

along the entire bonded portion of the interface.

In the boundary element model, each region again utilizes eighty-six quadratic elements, as in the sector model for Problem 1. Weighted traction results shown in Figs. 9 and 10 indicate that very good accuracy is obtained, although for this bi-material problem the maximum interface error of approximately 0.2% is significantly greater than for the corresponding single material analysis. For the Dirichlet problem, the weighted boundary element tractions obtained at the crack tip are $s_1 = 1.0000$ and $s_2 = 1.0001$, thus providing four digits of accuracy. Meanwhile, for the Neumann problem, $s_1 = 0.99926$ and $s_2 = 0.99910$ at the crack tip.

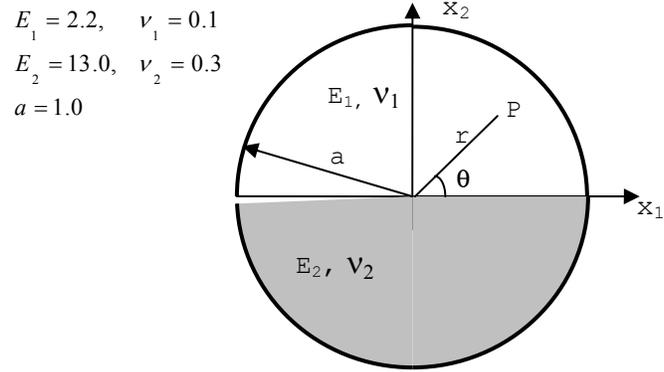

**Fig. 8.** Problem 3 bi-material sector definition

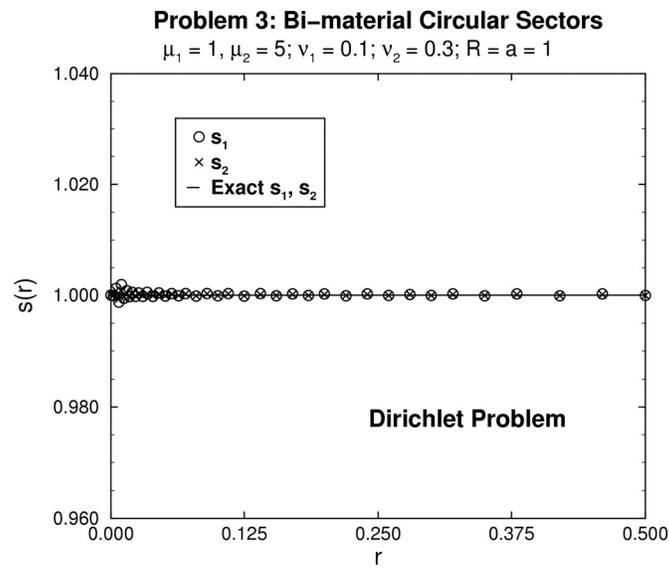

**Fig. 9.** Problem 3 bi-material sector results for Dirichlet boundary conditions

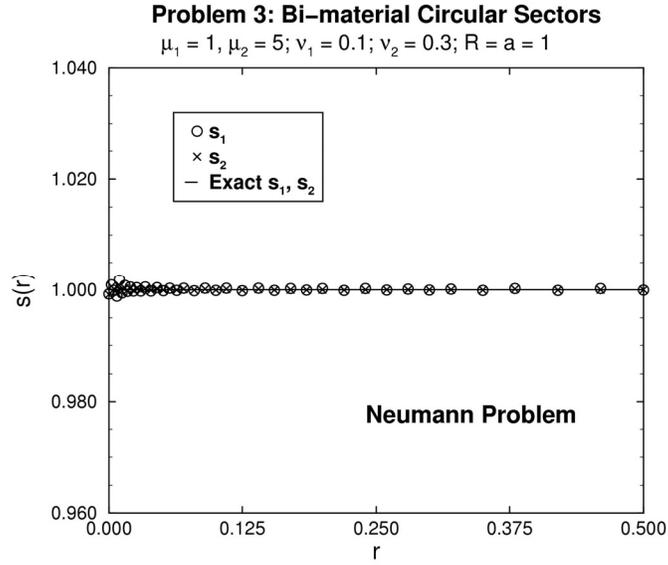

**Fig. 10.** Problem 3 bi-material sector results for Neumann boundary conditions

*6.5. Problem 4: Bonded elastic rectangular plates*

As a fourth example, we consider two bi-material plate problems analyzed in the literature. The first case involves the problem studied by Raveendra and Banerjee (1991), as defined in Fig. 11. The weighted traction results from a boundary element convergence study are displayed in Fig. 12. Here the coarse mesh utilizes sixty-seven quadratic elements for each domain, while the fine mesh employs exactly twice as many elements. Notice that very good convergence of the boundary element solutions is obtained. Converting these results to stress intensity factors using (58) for weighted tractions and (13) for crack opening displacements, we obtain the results shown in Table 3. The Raveendra and Banerjee (1991) solution was obtained using a boundary element formulation that employed a standard $r^{-1/2}$ traction singular approach, along with (13) to estimate the stress intensity factor from the resulting crack opening displacements.

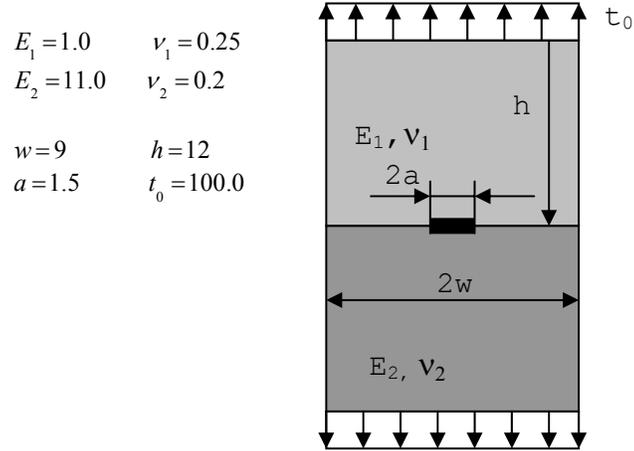

**Fig. 11.** Problem 4 bi-material plate definition (Raveendra and Banerjee, 1991)

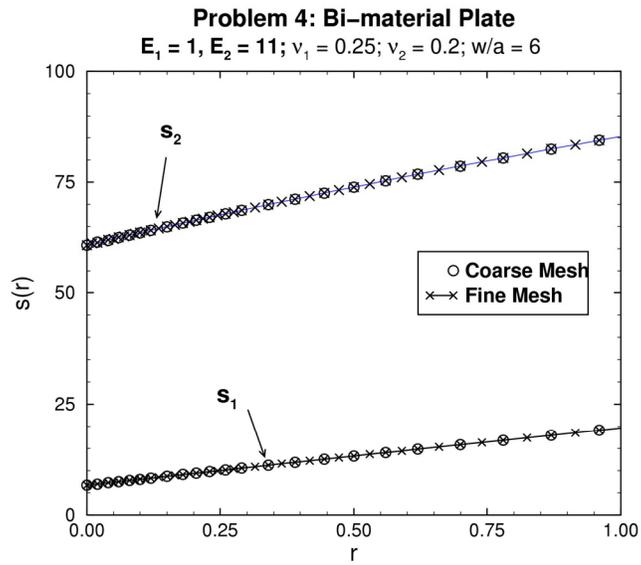

**Fig. 12.** Problem 4 weighted traction convergence study results

**Table 3.** Problem 4 stress intensity factors for the Raveendra and Banerjee problem

|  | $K_1$ | $K_2$ |
|---|---|---|
| Raveendra and Banerjee (1991) | 211.74 | 16.67 |
| BEM (coarse mesh - weighted tractions) | 213.61 | 14.00 |
| BEM (fine mesh - weighted tractions) | 213.74 | 14.06 |
| BEM (coarse mesh - crack opening displacements) | 213.31 | 14.46 |
| BEM (fine mesh - crack opening displacements) | 213.64 | 14.37 |

For the second case, we consider the central crack bi-material plate example presented by initially by Yuuki and Cho (1989). In this problem, $\mu_1 = 10$, $\nu_1 = 0.3$, $\mu_2 = 1$, $\nu_2 = 0.3$, $2a = w$ and $2w = h$. Results are presented in Table 4 in comparison with the boundary element solutions by Yuuki and Cho (1989) and Lim *et al.* (2002). Here the Lim *et al.* (2002) non-dimensional stress intensity factor $F$ is used for comparison, where

$$F = \frac{Ka^{-1/2+i\varepsilon}}{(2\pi)^{1/2}} \tag{91}$$

Notice that convergence of the present weighted traction boundary element formulation is quite good with both mesh refinement and overall problem scale (*i.e.*, variation of $a$). The converged magnitude of $F$ is within 4% of the previously reported solutions, although the ratio $F_2/F_1$ is significantly different. As a further comparison, the non-dimensional stress intensity factor obtained using the Raveendra and Banerjee (1991) approach is included in the table for two different levels of mesh refinement.

**Table 4.** Problem 4 non-dimensional stress intensity factors for the Yuuki and Cho problem

|  | $a$ | $m$ | $F_1$ | $F_2$ |
|---|---|---|---|---|
| Yuuki and Cho (1989) | - | 98 | 0.822 | -0.0974 |
| Lim *et al.* (2002) | - | 14 | 0.816 | -0.102 |
| BEM (mesh A - weighted tractions) | 0.25 | 14 | 0.787 | -0.052 |
| BEM (mesh B - weighted tractions) | 0.25 | 28 | 0.795 | -0.044 |
| BEM (mesh C - weighted tractions) | 0.25 | 56 | 0.796 | -0.044 |
| BEM (mesh D - weighted tractions) | 0.25 | 112 | 0.797 | -0.044 |
| BEM (mesh D - weighted tractions) | 0.50 | 112 | 0.797 | -0.044 |
| BEM (mesh D - weighted tractions) | 1.00 | 112 | 0.798 | -0.044 |
| BEM (mesh E - weighted tractions) | 0.25 | 224 | 0.797 | -0.045 |
| Raveendra and Banerjee approach | 0.25 | 56 | 0.791 | -0.053 |
| Raveendra and Banerjee approach | 0.25 | 112 | 0.792 | -0.052 |

*6.6. Problem 5: Epoxy-metal butt joint*

For the final example, we consider the problems of epoxy-steel and epoxy-aluminum butt joints loaded in tension. Figure 13 depicts the case with steel adherends, based upon physical experiments conducted by Reedy and Guess (1993, 1997) on cylindrical specimens. In their work, Reedy and Guess studied the dependence of joint strength on bond thickness and found that thinner adhesive layers provided enhanced strength, as is well known from common experience. More importantly, their quantitative analysis led to the discovery that the strength-thickness variation followed power law behavior with an exponent close to that of the free-edge elastic bi-material corner singularity. Thus, for the epoxy-steel joint, the analytical exponent is $\lambda_c - 1 \approx -0.30$, where $\lambda_c$ represents the eigenvalue obtained from a bi-material wedge (Bogy, 1971). The corresponding exponent for the epoxy-aluminum joint is $\lambda_c - 1 \approx -0.27$. In previous work, we developed a weighted traction axisymmetric boundary element formulation for fully bonded bimaterial interfaces and then confirmed that the generalized stress intensity factor (or weighted traction $t^\phi$) versus bond thickness for these epoxy-steel butt joints does indeed follow power laws with exponent $1 - \lambda_c$ (Dargush and Hadjesfandiari, 2004).

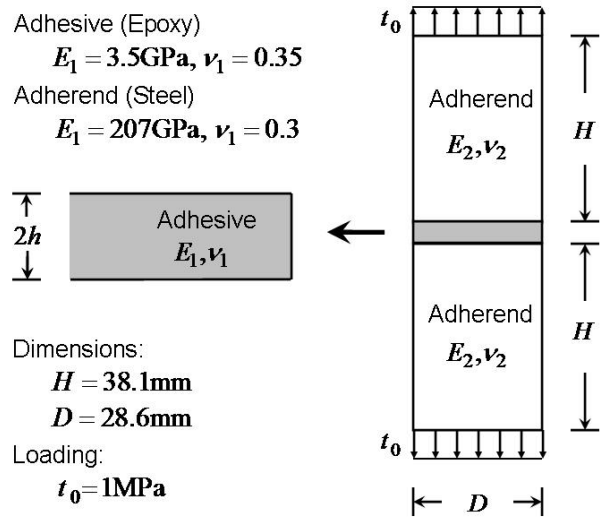

**Fig. 13.** Epoxy-steel butt joint problem definition (Reedy and Guess, 1993)

While this approach successfully approximates the characteristic power law behavior of the bonded joints, there are some concerns. First, the generalized stress intensity factor (or similarly, $t^\phi$) does not relate directly to energy release rates. Consequently, it is difficult to connect $t^\phi$ to the initiation or propagation of an interface crack that may lead to failure. Secondly, the critical amplitude of the generalized stress intensity factor (or $t_{cr}^{\phi}$) depends upon the two interfacing materials and also on the local geometry at the free edge. For example, if the angle between the interface and free edge were to change from $\pi/2$, then the singularity exponent $\lambda_c - 1$ would change, along with the critical amplitude $t_{cr}^{\phi}$. Thus, physical testing would be required to establish $t_{cr}^{\phi}$ for each geometric configuration.

As an alternative approach, here we postulate small edge cracks on the interface and then investigate the scaling behavior of the complex stress intensity factor (or weighted tractions $s(0)$) as a function of the thickness $2h$ of the adhesive, under conditions of plane strain. For the boundary element analysis, we impose quarter symmetry and use two levels of mesh refinement to examine convergence for small edge cracks with length $a = 0.010$ mm. The coarse mesh employs sixty-four quadratic elements to represent the epoxy half-layer, along with sixty-eight elements for the adherend. The fine mesh includes exactly twice as many elements. For the epoxy elastic material properties, $E = 3.5$ GPa and $\nu = 0.35$, while for steel $E = 207$ GPa and $\nu = 0.30$. As a result, from (5), $\varepsilon = 0.07182$. Meanwhile, for aluminum, $E = 69$ GPa and $\nu = 0.33$, which leads to $\varepsilon = 0.06687$.

Weighted tractions obtained from the present boundary element formulation at the crack tip are shown in Fig. 14a for the epoxy-steel case and in Fig. 14b for epoxy-aluminum connections. In each case, results from the coarse mesh and fine mesh are indistinguishable from each other in the scale of these figures. Interestingly, we find power law behavior for the variation of stress intensity factor (or weighted tractions) versus bond thickness with exponents of approximately 0.33 for steel adherends and 0.29 for the epoxy-aluminum butt joint.

Meanwhile, Figures 15a and b provide the joint strength versus bond thickness data obtained by Reedy and Guess (1997) in their physical experiments. Although there is significant variability in the experiments, the epoxy layer thickness effect on joint strength is clearly evident. Included in the figures are power law regressions that indicate approximate slopes of $-0.35$ and $-0.26$ for the steel and aluminum adherends, respectively. If we assume that failure occurs at a given critical value of the weighted tractions (or complex stress intensity factor), then the experimental results are reasonably consistent with the present boundary element solutions, which would provide slopes of $-0.33$ for epoxy-steel butt joints and $-0.29$ for the epoxy-aluminum interfaces. This naturally suggests that the present boundary element formulation may useful in estimating the scaling behavior of such joints.

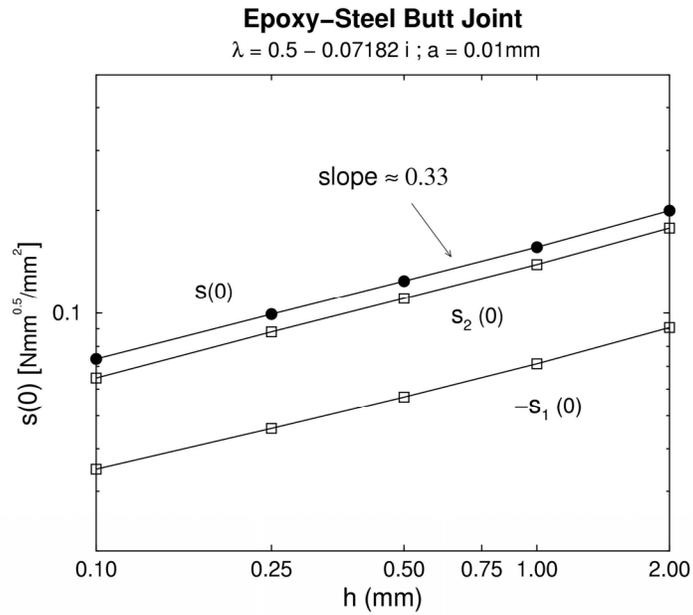

**Fig. 14.** BEM weighted tractions for edge crack $a = 0.010$ mm (a) Epoxy-steel butt joint

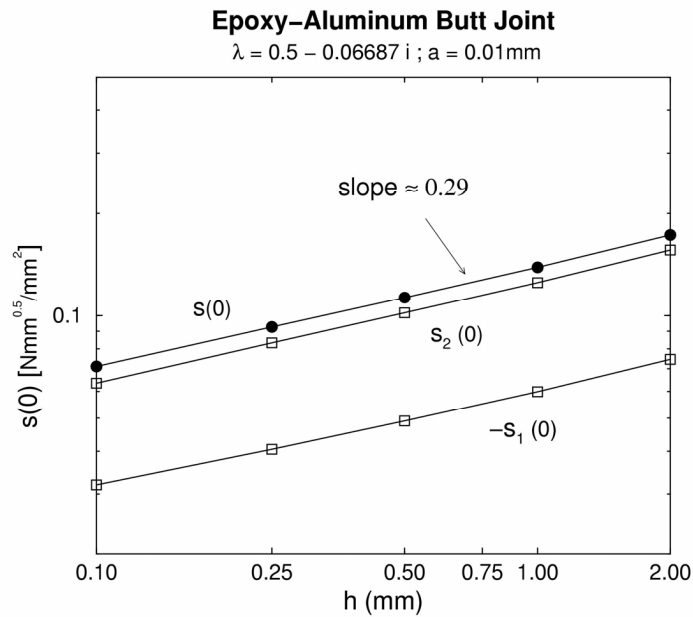

**Fig. 14.** BEM weighted tractions for edge crack $a = 0.010$ mm (b) Epoxy-aluminum butt joint

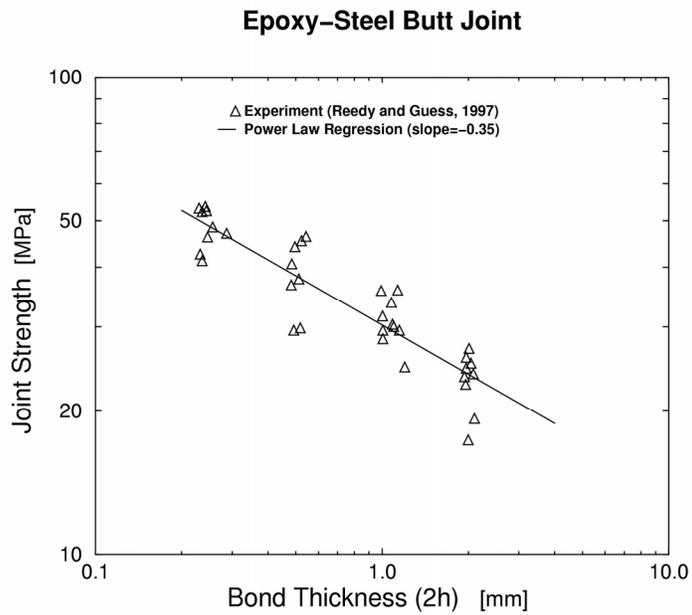

**Fig. 15.** Butt joint experimental results (a) Epoxy-steel specimens (Reedy and Guess, 1997)

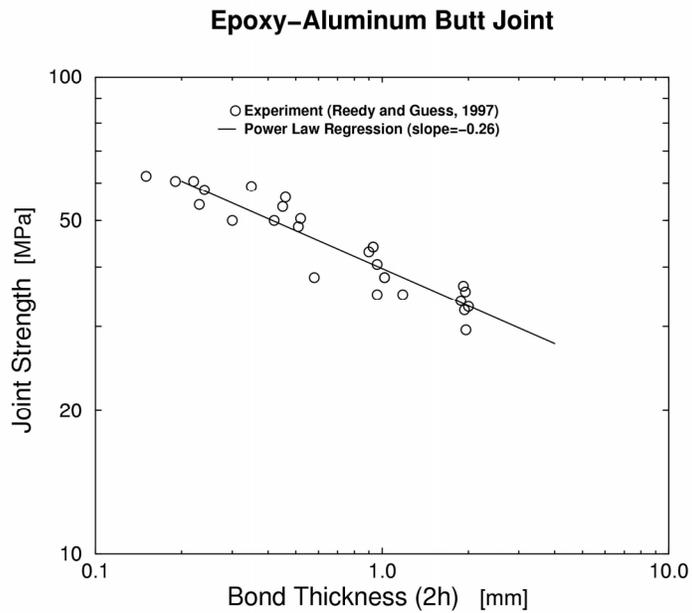

**Fig. 15.** Butt joint experimental results (b) Epoxy-aluminum specimens (Reedy and Guess, 1997)

## 7. Conclusions

In this paper, a boundary element formulation and numerical implementation is developed to determine the complex stress intensity factors for cracks on bi-material interfaces. The boundary element method formulated previously in (Raveendra and Banerjee, 1991) ignores the sub-element level log-periodic oscillations and therefore provides only approximations to the linear elastic bi-material fracture mechanics problem. The more recent work by Lim et al. (2002) considers only partially all of the kernel singularities. On the other hand, the present methodology addresses these contributions in a more rigorous manner and provides highly accurate numerical solutions to the underlying mathematical problem, as demonstrated through a number of rigorous numerical examples. A final application to strength analysis of epoxy-metal butt joints suggests that the method may be useful in estimating the scaling relations that result from the presence of bi-material interfaces. However, we must caution that the present formulation is, of course, an approximation and, for example, the linear theory does entail small-scale interpenetrations of the crack surfaces near the tip.

## References

Banerjee, P.K., 1994. The Boundary Element Methods in Engineering. McGraw-Hill, London.

Barsoum, R.S., 1974. Application of quadratic isoparametric finite-elements in linear fracture mechanics. *Int. J. Fract.*, 10, 603-605.

Blandford, G.E., Ingraffea, A.R., Liggett, J.A., 1981. Two-dimensional stress intensity factor computations using the boundary element method. *Int. J. Numer. Meth. Engrg.*, 17, 387-404.

Bogy, D.B., 1971. Two edge-bonded elastic wedges of different materials and angles under surface tractions. *J. Appl. Mech.*, ASME, 38, 377-386.

Comninou, M., 1977. The interface crack. *J. Appl. Mech.*, 44, 631-636.

Cruse, T.A., 1978. 2-dimensional BIE fracture mechanics analysis. *Appl. Math. Modelling*, 2, 287-293.

Cruse, T.A., 1996. BIE fracture mechanics analysis: 25 years of developments. *Comp. Mech.*, 18, 1-11.

Dargush, G.F., Hadjesfandiari, A.R. 2004. Generalized stress intensity factors for strength analysis of bi-material interfaces. *Mech. Adv. Mater. Struct.*, 11, 1-15.

England, A.H., 1965. A crack between dissimilar media. *J. Appl. Mech.*, 32, 400-402.

Erdogan, F., 1965. Stress distribution in bonded dissimilar materials with cracks. *J. Appl. Mech.*, 32, 403-410.

Golub, G.H., Welsch, J.H., 1969. Calculation of Gauss quadrature rules. *Math. Comput.*, 23, 221-230.

Hutchinson, J.W., Mear, M., Rice, J.R., 1987. Crack paralleling an interface between dissimilar materials. *J. Appl. Mech.*, 54, 828-832.

Lee, K.Y., Choi, H.J., 1988. Boundary element analysis of stress intensity factors for bi-material interface cracks. *Engrg. Fract. Mech.*, 29, 461-472.

Lee, S.S., 1996. Boundary element evaluation of stress intensity factors for interface edge cracks in a unidirectional composite. *Engrg. Fract. Mech.*, 55, 1-6.

Lim, K.M., Lee, K.H., Tay, A.A.O., Zhou, W., 2002. A new variable-order singular boundary element for stress analysis. *Int. J. Numer. Meth. Engrg.*, 55, 293–316.

Lin, K.Y., Mar, J.W., 1976. Finite element analysis of stress intensity factors for cracks at a bi-material interface. *Int. J. Fract.*, 12, 521-531.

Muskhelishvili, N.I., 1953. Some Basic Problems of the Mathematical Theory of Elasticity. P. Noordhoff, Groningen, The Netherlands.

Press, W.H., Teukolsky, S.A., Vetterling, W.T., Flannery, B.P., 1992. Numerical Recipes in Fortran. The Art of Scientific Computing. Cambridge University Press, Cambridge.

Reedy, Jr., E.D., Guess, T.R. 1993. Comparison of butt tensile strength data with interface corner stress intensity factor prediction. *Int. J. Solids Struct.*, 30, 2929-2936.

Reedy, Jr., E.D., Guess, T.R. 1997. Interface corner failure analysis of joint strength: Effect of adherend stiffness. *Int. J. Fract.*, 88, 305-314.

Raveendra, S.T., Banerjee, P.K., 1991. Computation of stress intensity factors for interfacial cracks. *Engrg. Fract. Mech.*, 40, 89-103.

Rice, J.R., 1988. Elastic fracture mechanics concepts for interfacial cracks. *J. Appl. Mech.*, 55, 98-103.

Rice, J.R., Sih, G.C., 1965. Plane problems of cracks in dissimilar media. *J. Appl. Mech.*, 32, 418-423.

Sack, R.A., Donovan, A.F., 1972. Algorithm of Gaussian quadrature given modified moments. *Numer. Math.*, 18, 465-478.

Sih, G.C., Rice, J.R., 1964. The bending of plates of dissimilar materials with cracks. *J. Appl. Mech.*, 31, 477-482.

Sornette, D., 1998. Discrete-scale invariance and complex dimensions. *Phys. Reports*, 297, 239-270.

Watson, J.O., 1995. Singular boundary elements for the analysis of cracks in plane-strain. *Int. J. Numer. Meth. Engrg.*, 38, 2389-2411.

Wheeler, J.C., 1974. Modified moments and Gaussian quadratures. *Rocky Mtn. J. Math.*, 4, 287-296.

Williams, M.L., 1959. The stresses around a fault or crack in dissimilar media. *Bull. Seismol. Soc. Amer.*, 49, 199-204.

Yuuki R., Cho S.B., 1989. Efficient boundary element analysis of stress intensity factors for interface cracks in dissimilar materials. *Engrg. Fract. Mech.*, 34, 179–188.

Zhou, W., Lim, K.M., Lee, K.H., Tay, A.A.O., 2005. A new variable-order singular boundary element for calculating stress intensity factors in three-dimensional elasticity problems. *Int. J. Solids Struct.*, 42, 159-185.